\documentclass{rspublic}
\usepackage{bm}
\usepackage{epsfig}

\newcommand{\osum}
   {\mathrel{\rlap{\raise-.5pt\hbox{\Large{\hspace{6.6pt}\texttt{o}}}}
{\hbox{$\displaystyle\sum_{<\alpha\beta>}$}}}}
\newcommand{\smallosum}
   {\mathrel{\rlap{\raise-.2pt\hbox{\large{\hspace{1.2pt}\texttt{o}}}}
{\hbox{$\sum$}}}}
\newcommand{\osumtrait}
   {\mathrel{\rlap{\raise-.5pt\hbox{\Large{\hspace{4.6pt}\texttt{o}}}}
{\hbox{$\displaystyle\sum_{\alpha\to\beta}^{(\partial B)}$}}}}
\newcommand{\cp}{\text{c.\,p.}}

\begin{document}
\title[Up to next-nearest neighbour elasticity of triangular
lattices]{General up to next-nearest neighbour elasticity of triangular
lattices\\ in three dimensions}

\author[C. Dubus, K. Sekimoto, and J.-B. Fournier]{Cyril Dubus$^{1,2}$,
Ken Sekimoto$^{1,2}$, and Jean-Baptiste Fournier$^{1,2}$}

\affiliation{$^1$Laboratoire Mati\`ere et Syst\`emes Complexes, UMR 7057 CNRS \&
Universit\'e Paris~7, 2 place Jussieu, F-75251 Paris cedex 05, France. \\ 
$^2$Laboratoire de Physico-Chimie
Th\'eorique, UMR 7083 CNRS, ESPCI, 10 rue Vauquelin, F-75231
Paris cedex 05, France.}  

\label{first page}
\maketitle

\begin{abstract}{triangular lattice, lattice elasticity, non-central
forces,  cytoskeleton} We  establish the  most  general form  of the
discrete elasticity of  a 2D  triangular lattice embedded  in three
dimensions, taking into  account up  to next-nearest  neighbour
interactions.  Besides crystalline system,  this  is  relevant  to
biological  physics  (e.g.,  red  blood  cell cytoskeleton) and soft
matter (e.g., percolating gels, etc.). In order to correctly impose  the
rotational  invariance  of the  bulk  terms, it  turns  out to  be
necessary to take into account explicitly the elasticity  associated
with the vertices located at the edges of the lattice.  We find  that
some  terms that  were suspected in  the litterature  to violate
rotational symmetry are in fact admissible. 
\end{abstract}

\section{Introduction}

Since Born and von K\'arm\'an, the construction of microscopic elastic
models of crystals has formed an important part of solid-state physics,
aimed to determine elastic constants and lattice dynamical properties
(Born \& von K\'arm\'an 1912; Born 1914).  Imposing global translational
and rotational invariance is a delicate matter, although the 
required relationship are well established (Born \& Huang 1954).  In fact,
inconsistencies associated with rotational invariance in the earlier
models were pointed out by Lax (1965) and Keating (1966$a,b$).  
More recently, the 
importance of triangular lattices has emerged with the study of
soft matter systems such as percolating gels (Feng \& Sen 1984; Schwartz
\textit{et al.} 1985; Arbabi \&
Sahimi 1993), membrane with crystalline order (Kantor \& Nelson 1987; Seung \&
Nelson 1988; Gompper \& Kroll 1997), biological membranes (Discher \textit{et
  al}.\ 1997; Coughlin \& Stamenovic 2003), red blood cell
cytoskeleton (Saxton 1990; Lim \textit{et al}.\ 2002; Marcelli \textit{et al.}
2005), and finite element modeling (Gusev 2004).

In this paper, we consider a triangular lattice embedded in three
dimensions and we determine the most general expression of the harmonic
elastic energy up to next-nearest neighbour interactions.  We formulate
a systematic procedure to generate all the energy terms allowed by
symmetry. This procedure could be applied to other type of lattices. 
We show that in order to correctly ensure rotational
invariance, it is necessary to consider the boundary of the lattice, as
first suggested by Lax (1965).
 Our central result, equation~(\ref{FK}),
gives \textit{all} the allowed energy terms, with no particular connection to
mechanical elements such as springs, elastic wedges, etc. This contrasts with
the above cited papers that postulate energy terms in a non exhaustive
way. In particular, the most general energy contains a
non-central nearest neighbour term (with coefficient $K_t$), which was
believed to violate rotational invariance in percolating gels (Feng \& Sen
1984) on the grounds of Keating's argumentation (1966$a,b$).
The situation is actually delicate since in a later paper, 
Keating (1968) restricted the generality of some assertions of its earlier
papers. 

Our paper is organised as follows. In \S2 we formulate
the elastic energy of triangular lattices.  In \S2\,$a$, we write down
all the scalar energy terms quadratic in the displacements of the
lattice vertices. In \S2\,$b$, we retain only the 
terms satisfying the six-fold
symmetry of an infinite triangular lattice. In \S2\,$c$, we consider also the
edges: we study the 
symmetries relative to them and we retain the allowed edge terms.
In \S2\,$d$, we impose global translational and rotational invariance, 
taking explicitely into account the reduced symmetry of the boundary,
and we retain the allowed terms.  In \S2\,$e$, we collect all the
contributions and we extract the resulting bulk terms.  In \S3, we
analyse and discuss the results obtained in \S2.  In \S3\,$a$, we give a
possible interpretation of our energy in terms of mechanical elements.
In \S3\,$b$, we rewrite the energy into the generic expression
(\ref{FK}), which constitutes our central result.  In \S3\,$c$, we
establish the connection between our model and its continuum version.
Finally, in \S4, we give a brief summary and we detail the origin of the
rotational invariance problem.

\section{Discrete elasticity model}

\subsection{General definitions}

\begin{figure}
\centerline{\epsfig{file=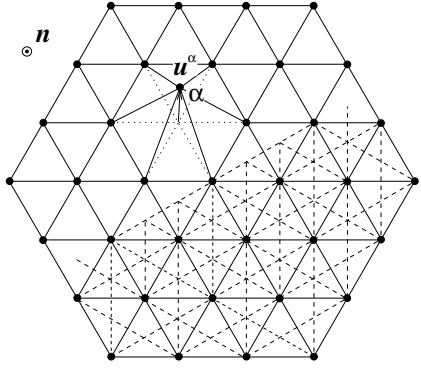,width=5.5cm}}
\caption{Triangular elastic lattice with the vertex $\alpha$ displaced
by $\bm{u}^\alpha$. Nearest neighbours are indicated by plain lines and
next-nearest neighbours by dashed lines (for clarity, the latters
are drawn on one half of the figure only). The edge length is $N\ell$
with $N=3$.}
\label{notations}
\end{figure}

Let us consider a bounded triangular lattice. In its undistorted state,
we assume that it lies in a plane $\mathcal{P}$ and that its vertices
(i.e.\ its sites) are evenly separated by a distance~$\ell$
(figure~\ref{notations}). We denote by $\bm{r}^\alpha\in\mathcal{P}$ the
position of a generic vertex~$\alpha$. We also assume that the elastic
properties of the lattice are symmetric with respect to~$\mathcal{P}$. 

In an arbitrary distorted state, $\bm{r}^\alpha$ becomes
$\bm{r}^\alpha+\bm{u}^\alpha$, where $\bm{u}^\alpha$ is a
three-dimensional displacement. The most general quadratic expansion of
the lattice elastic energy, $F$, is given by
\begin{equation}
\label{Fgen}
F\left(\left\{\bm{u}^\alpha\right\}\right)=
\sum_{\alpha}\bm{u}^\alpha\cdot\bm{S}^\alpha
\cdot\bm{u}^\alpha+\sum_{\alpha}\sum_{\beta\neq\alpha}\bm{u}^\alpha\cdot
\bm{A}^{\alpha\beta}\cdot \bm{u}^\beta\,.
\end{equation}
In this expression, we have separated the terms coupling identical
vertices 
from those coupling different vertices. We have excluded the terms linear in
$\bm{u}^\alpha$, because they must vanish for the undistorted state to be a minimum of
the energy. Since $F$ is scalar, the tensors $\bm{S}^\alpha$ and
$\bm{A}^{\alpha\beta}$ must satisfy
$(\bm{S}^\alpha)^\mathrm{t}=\bm{S}^\alpha$ and
$(\bm{A}^{\alpha\beta})^\mathrm{t}=\bm{A}^{\beta\alpha}$, where the
superscript t means transpose. Note that in general
$\bm{A}^{\alpha\beta}\neq\bm{A}^{\beta\alpha}$. 

We fix the \textit{interaction range} by assuming that all
the $\bm{A}^{\alpha\beta}$ vanish except when $\alpha$ and
$\beta$ are nearest or next-nearest neighbours.

\subsection{Bulk terms}

To begin with, we consider the \textit{bulk} vertices, i.e. the vertices
that are far away enough from the boundary (this point will be precised in
\S2$\,c$) 
so that the triangular
symmetry may be assumed.  
In the bulk the elastic properties are assumed to be spatially
homogeneous.
Let $\bm{n}$ be a unitary vector normal to
$\mathcal{P}$. Let ($\bm{t}_1$, $\bm{t}_2$) be two three-dimensional
vectors forming 
a \textit{local}
orthonormal basis in $\mathcal{P}$, allowed to vary from site to site.
The tensors $\bm{S}^\alpha$ and $\bm{A}^{\alpha\beta}$ can be decomposed
into the complete basis $\mathcal{B}=(\bm{t}_1\otimes\bm{t}_1,\,
\bm{t}_1\otimes\bm{t}_2,\, \bm{t}_2\otimes\bm{t}_1,\,
\bm{t}_2\otimes\bm{t}_2,\, \bm{n}\otimes\bm{n})$, where the symbol $\otimes$
represent the tensorial product. Elements linear in $\bm{n}$, such as
$\bm{t}_1\otimes\bm{n}$, have been excluded, since 
the undistorted lattice is symmetric with respect to $\mathcal{P}$.

To construct $\bm{S}^\alpha$, we take $\bm{t}_1$
along any one of the six nearest-neighbour directions of the undistorted lattice, and
$\bm{t}_2=\bm{n}\times\bm{t}_1$.  Because of the triangular symmetry, each
term in the decomposition of $\bm{S}^\alpha$ must be even in $\bm{t}_1$
and in $\bm{t}_2$, and the
coefficients of $\bm{t}_1\otimes\bm{t}_1$ and $\bm{t}_2\otimes\bm{t}_2$
must be equal (this condition is necessary for $\bm{S}^\alpha$ to be
invariant under an in-plane $\frac{\pi}{3}$ rotation). Hence, 
\begin{equation}
\label{S}
\bm{S}^\alpha=s_1\,\bm{I_2}+\tilde{s}_1\,\bm{n}\otimes\bm{n},
\end{equation}
where $\bm{I_2}=\bm{t}_1\otimes\bm{t}_1+\bm{t}_2\otimes\bm{t}_2$ 
is the identity tensor in the plane $\mathcal{P}$.
By virtue of the lattice homogeneity, the coefficients $s_1$ and
$\tilde{s}_1$ should not depend on $\alpha$.

We now construct $\bm{A}^{\alpha\beta}$. Let
$\bm{t}^{\alpha\beta}$ be the unit vector in the direction going from
$\bm{r}^\alpha$ to $\bm{r}^\beta$. We take
$\bm{t}_1=\bm{t}^{\alpha\beta}$ and $\bm{t}_2=\bm{n}\times\bm{t}_1$.
Without loss of generality the decomposition of $\bm{A}^{\alpha\beta}$
over $\mathcal{B}$ can be rewritten as
\begin{equation}
\label{Ageneral}
\bm{A}^{\alpha\beta}=\frac{a_i}{2}\,\bm{I_2}
+\frac{\tilde{a}_i}{2}\,\bm{n}\otimes\bm{n}
+\frac{\bar{a}_i}{2}\,\bm{t}_1\otimes\bm{t}_1
+\frac{\bar{\bar{a}}_i}{2}\,
(\bm{t}_1\otimes\bm{t}_2+\bm{t}_2\otimes\bm{t}_1)+\frac{a^*_i}{2}
(\bm{t}_1\otimes\bm{t}_2-\bm{t}_2\otimes\bm{t}_1).
\end{equation}
By virtue of the lattice homogeneity, the coefficients 
$\{a_i,\tilde{a}_i,\ldots\}$ take only two
sets of values: either $\{a_1,\tilde{a}_1,\ldots\}$ if $\alpha$ and
$\beta$ are nearest neighbours, or $\{a_2,\tilde{a}_2,\ldots\}$ if
$\alpha$ and $\beta$ are next-nearest neighbours.  The
in-plane mirror symmetry with respect to $\bm{t}^{\alpha\beta}$ requires
that $\bm{A}^{\alpha\beta}$ must be even in $\bm{t}_2$, which leads to
\begin{equation}
\label{Ageneral-bis}
\bar{\bar{a}}_i=a^*_i=0.
\end{equation}

\subsubsection{Collecting bulk terms} 
\begin{figure}
\centerline{\epsfig{file=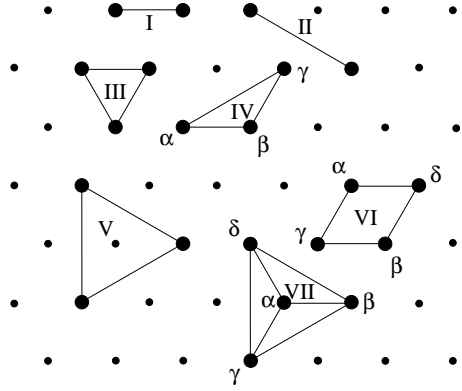,width=6cm}}
\caption{The seven types of neighbour clusters as indicated by the
large dots (the lines are only guides for the eyes).}
\label{clusters}
\end{figure}

We now anticipate that we shall clearly
distinguish between the $\bm{S}^\alpha$ and $\bm{A}^{\alpha\beta}$
belonging to the bulk and those belonging to the edges. Denoting by
$F_{\text{bulk}}$ the part of the energy $F$ which involves 
only the \textit{bulk}
terms, we arrive, from equation~(\ref{Fgen}), (\ref{S}), (\ref{Ageneral}) and
(\ref{Ageneral-bis}), at
\begin{eqnarray}
\label{Favantrans}
F_{\text{bulk}}\!\!&=&\!\!
\sum_{\alpha}^\text{(bulk)}\left[s_1(\bm{u}^\alpha_\perp)^2
+\tilde{s}_1(u^\alpha_n)^2\right] \nonumber
\\
\!\!&+&\!\!\sum_{<\alpha\beta>}^\text{(bulk)}\sum_{i={1}}^{2}
\delta^{\alpha\beta}_i
\left[a_i\,\bm{u}^\alpha_\perp\cdot\bm{u}^\beta_\perp
+\tilde{a}_i\,u^\alpha_n\,u^\beta_n
+\bar{a}_i\,(\bm{u}^\alpha_\perp\cdot\bm{t}^{\alpha\beta})\,
(\bm{u}^\beta_\perp\cdot\bm{t}^{\alpha\beta})\right]\!,
\end{eqnarray}
where $u_n^\alpha\equiv \bm{u}^\alpha\cdot\bm{n}$ 
denotes the normal component of
$\bm{u}^\alpha$ and 
$\bm{u}_\perp^\alpha\equiv\bm{u}^\alpha-u_n^\alpha\bm{n}$
the projection of
$\bm{u}^\alpha$ on $\mathcal{P}$. The symbol $\sum_{<\alpha\beta>}$
indicates the summation over both nearest and next-nearest neighbour
pairs of vertices. As already mentioned, we assign to the former the
suffix $i=1$ and to the latter the suffix $i=2$ 
(see figure~\ref{clusters} where these
numbers are indicated as I and II).  The function
$\delta^{\alpha\beta}_i$ takes the value $1$ if the pair \{$\alpha$,
$\beta$\} is of type $i$ and zero otherwise. We shall hereafter call
such functions \textit{indicatrix functions}.

\subsection{Edge terms}

With arbitrary coefficients $s_1,\tilde{s}_1,a_i,\tilde{a}_i$ and
$\bar{a}_i$ ($i=1,2$), the energy form (\ref{Favantrans}) is not
invariant under a global translation or rotation of the lattice. 
In order to correctly ensure rotational invariance, it is necessary to take
into account the \textit{edges} of the lattice. 
This seems counterintuitive, because one expects the contribution of the edges
to be subdominant with respect to the expensive contribution of the
bulk. However, as it will more precisely be shown in \S 4, this is not the
case, because in a rotation the displacements of the edges are proportional to
the size of the system.

\subsubsection{Neighbour clusters and vertices types}
\label{subsubsec:type}

Since our interest lies in the bulk terms, we choose the simple shape of a
regular hexagon of edge length $N\ell$ (figure~\ref{notations}) as our system.
Consistently with the energy expansion,
 the specificity of the edges must
be taken into account up to the next-nearest neighbour
distance. For this purpose, we group the vertices into 
\textit{neighbour clusters}
in which all the inter-distances are at most equal to the next-nearest
neighbour distance ($\ell\sqrt{3}$). There are seven types of such
clusters, which we label from I to VII (see figure~\ref{clusters}). A
vertex is said to participate in a neighbour cluster of type Y if it 
can be grouped with its neighbour vertices into a neighbour cluster of type Y.
Note that vertices near the edges 
participate in a smaller number of neighbour clusters than those in the bulk,
because of their lacking neighbours.

\begin{figure}
\centerline{\epsfig{file=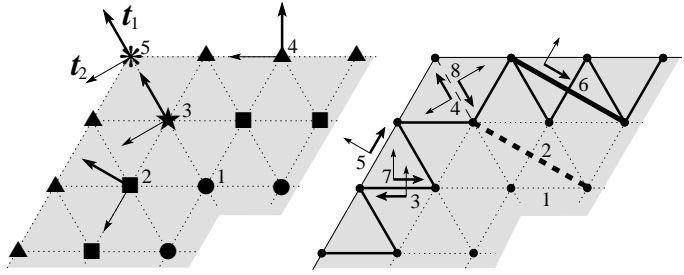,width=9cm}}
\caption{The 5 types of vertices (left) and the 8 types of vertices
doublets (right) together with their local basis $\bm{t}_1$, $\bm{t}_2$
(the former being in thick arrows). Only the basis relative to the
edges are indicated. For the doublet $(\alpha,\beta)$ of types 3,4,7 and 8,
the begininng (resp.\ end) of the thick arrow corresponds to the first 
(resp.\ second) component $\alpha$ (resp. $\beta$) of the doublet.}
\label{bords}
\end{figure}

To each vertex $\alpha$ we now assign a \textit{type} depending on
which neighbour clusters it participates in: two vertices are assigned the
same type if, and only if, their positions within the set of neighbour cluster
in 
which they participate are indistinguishable. Close inspection shows that
the edge vertices are classified into five types, 
labeled from 1 to 5, as shown in
figure~\ref{bords}. The vertices of type 1 are the bulk vertices.

We apply a similar procedure to assign a type to the oriented 
nearest and next-nearest neighbour vertex pairs, which we call
 \textit{vertex doublets}.
We distinguish the vertex doublets  
$(\alpha,\beta)$ from $(\beta,\alpha)$ 
since generally
$\bm{A}^{\alpha\beta}\neq\bm{A}^{\beta\alpha}$. Two vertex doublets
are assigned the same type if, and only if, their positions and
orientations within the set of neighbour clusters in which they participate are
indistinguishable. We thus determine eight types of vertex doublets,
labeled from 1 to 8 (figure~\ref{bords}). The vertex doublets of type 1
are the bulk nearest-neighbour vertex doublets and those of type 2 are the
bulk next-nearest neighbour vertex doublets.
We also define
 types for the vertex \textit{pairs} in the following 
way. When the two doublets associated with a given pair are of the same
type, the pair is assigned that type (this holds for the types
$1,2,5,6$) while the pairs corresponding to doublet types 3 and 7
are assigned the type 3 and those corresponding to doublet types 4 and 8
are assigned the type 4.

\subsubsection{Symmetry allowed edge terms}

In order to take into account the local symmetry up to the next-nearest
neighbour distance, we consider the \textit{symmetries of the set of neighbour
  clusters} in which
the vertices or vertex doublets participate. Let us
first examine the coefficients of the tensors $\bm{S}^\alpha$ for
vertices of type 2 to 5 (edge vertices). Since each one possesses an axis
of symmetry, we choose $\bm{t}_1$ along this axis and
$\bm{t}_2\perp\bm{t}_1$ as indicated in figure~\ref{bords}. Then, as
$\bm{S}^\alpha$ must be even in $\bm{t}_2$, we arrive at 
\begin{equation}
\bm{S}^\alpha=
s_i\,\bm{I_2}+\tilde{s}_i\,\bm{n}\otimes\bm{n}+\bar{s}_i
\,\bm{\tau}^\alpha\otimes\bm{\tau}^\alpha, 
\qquad (i={2},\ldots,{5})
\end{equation} 
where $\bm{\tau}^\alpha$ may either $\bm{t}_1$ or $\bm{t}_2$. For
definiteness, we choose $\bm{\tau}^\alpha=\bm{t}_2$ for vertices of
type 2 or 4 and $\bm{\tau}^\alpha=\bm{t}_1$ for vertices of type 3 or 5.

We now construct the tensor $\bm{A}^{\alpha\beta}$ for the vertex doublets of type
3 to 8 (edge types). The most general expression for
$\bm{A}^{\alpha\beta}$ is given by an expression similar to
(\ref{Ageneral}):
\begin{eqnarray}
\label{A2}
\bm{A}^{\alpha\beta}\!\!&=&\!\!\frac{a_{\alpha\beta}}{2}\,\bm{I_2}
+\frac{\tilde{a}_{\alpha\beta}}{2}\,\bm{n}\otimes\bm{n}
+\frac{\bar{a}_{\alpha\beta}}{2}\,\bm{t}_1\otimes\bm{t}_1\nonumber\\
\!\!&+&\!\!\frac{\bar{\bar{a}}_{\alpha\beta}}{2}\,
(\bm{t}_1\otimes\bm{t}_2+\bm{t}_2\otimes\bm{t}_1)+
\frac{a^*_{\alpha\beta}}{2} (\bm{t}_1\otimes\bm{t}_2-
\bm{t}_2\otimes\bm{t}_1).
\end{eqnarray}
We choose $\bm{t}_1=\bm{t}^{\alpha\beta}$ and $\bm{t}_2$ as indicated in
figure~\ref{bords}. The vertex doublets of types 4 and 8 have a mirror
symmetry with respect to $\bm{t}_1$, hence $\bm{A}^{\alpha\beta}$ must
be even in $\bm{t}_2$, which implies
$\bar{\bar{a}}^{\alpha\beta}=a^{*\alpha\beta}=0$. Other relationships
can be obtained by comparing the coefficients of $\bm{A}^{\alpha\beta}$
and those of $\bm{A}^{\beta\alpha}$. First, note that
$(\bm{A}^{\alpha\beta})^\mathrm{t}=\bm{A}^{\beta\alpha}$ implies
$a^{\alpha\beta}=a^{\beta\alpha}$,
$\tilde{a}^{\alpha\beta}=\tilde{a}^{\beta\alpha}$,
$\bar{a}^{\alpha\beta}=\bar{a}^{\beta\alpha}$,
$\bar{\bar{a}}^{\alpha\beta}=-\bar{\bar{a}}^{\beta\alpha}$ and
$a^{*\alpha\beta}=a^{*\beta\alpha}$. Then, since the vertex doublets
of type 5 and 6 do not change type when exchanging $\alpha$ and $\beta$,
their coefficients must satisfy
$a^{\alpha\beta}=a^{\beta\alpha}$,
$\tilde{a}^{\alpha\beta}=\tilde{a}^{\beta\alpha}$,
$\bar{a}^{\alpha\beta}=\bar{a}^{\beta\alpha}$,
$\bar{\bar{a}}^{\alpha\beta}=\bar{\bar{a}}^{\beta\alpha}$ and
$a^{*\alpha\beta}=a^{*\beta\alpha}$. Therefore
$\bar{\bar{a}}^{\alpha\beta}=0$ for types 5 and 6. Finally, we find
\begin{equation}
\bm{A}^{\alpha\beta}=\frac{a_i}{2}\,\bm{I_2}
+\frac{\tilde{a}_i}{2}\,\bm{n}\otimes\bm{n}
+\frac{\bar{a}_i}{2}\,\bm{t}^{\alpha\beta}    
\otimes\bm{t}^{\alpha\beta}+\frac{a_i^{*}}{2} \,\bm{\epsilon}^{\alpha\beta}
+\frac{\bar{\bar{a}}_i}{2}\,\bm{\sigma}^{\alpha\beta},
\quad (i=3,\ldots,8)
\end{equation}
where
$\bm{\epsilon}^{\alpha\beta}=\bm{t}_1\otimes\bm{t}_2-\bm{t}_2\otimes\bm{t}_1$,
$\bm{\sigma}^{\alpha\beta}=\bm{t}_1\otimes\bm{t}_2+\bm{t}_2\otimes\bm{t}_1$
and $a_4^*=a_8^*=\bar{\bar{a}}_4=\bar{\bar{a}}_{8}=\bar{\bar{a}}_5=
\bar{\bar{a}}_6=0$. In the following, we define $\bm{\epsilon}$ as the
tensor that performs a counterclockwise $\frac{\pi}{2}$ rotation in the plane
$\mathcal{P}$; depending on the relative orientations of
$\bm{t}_1$ and $\bm{t}_2$, $\bm{\epsilon}^{\alpha\beta}$ is either equal
to $\bm{\epsilon}$ or to $-\bm{\epsilon}$. Note that in general, we
do find $\bm{A}^{\alpha\beta}\neq\bm{A}^{\beta\alpha}$, contrary to what
is the case in the bulk. 

\subsubsection{Collecting edge terms}
\label{toto}

The total elastic energy can now be written as
$F=F_\text{bulk}+F_\text{edge}$, with $F_\text{bulk}$ given by
(\ref{Favantrans}) and $F_\text{edge}$ obtained by collecting the edge
terms defined above. We regroup the terms
$\bm{u}^\alpha\cdot\bm{A}^{\alpha\beta}\cdot\bm{u}^{\beta}$ and
$\bm{u}^\beta\cdot\bm{A}^{\beta\alpha}\cdot\bm{u}^{\alpha}$ in order to
sum on the pairs and not on the doublets.
We arrive then at 
\begin{eqnarray}
\label{Fedgeavantrans}
F_{\text{edge}}\!\!&=&\!\!\sum_{\alpha}\sum_{i={2}}^{5}\delta^{\alpha}_i 
\left[s_i(\bm{u}^\alpha_\perp
)^2 +\tilde{s}_i(u^\alpha_n)^2
+\bar{s}_i(\bm{u}^\alpha_\perp \cdot \bm{\tau}^{\alpha})^2\right] \nonumber\\ 
&+&\!\!\sum_{<\alpha\beta>}\sum_{i={3}}^{6}\delta^{\alpha\beta}_i\left[a_i \, 
  \bm{u}^\alpha_\perp \cdot 
  \bm{u}^\beta_\perp 
  +\tilde{a}_i \, u^\alpha_n u^\beta_n
  +\bar{a}_i (\bm{u}_\perp^\alpha \cdot \bm{t}^{\alpha\beta})
  (\bm{u}_\perp^\beta \cdot \bm{t}^{\alpha\beta})\right] \nonumber \\
&+&\!\!\!\!\!\osum\!
\left[\delta^{\alpha\beta}_3\,\eta^{\alpha\beta}\,\bar{\bar{a}}_3 
  \,\bm{u}_\perp^\alpha \cdot \bm{\sigma}^{\alpha\beta} \cdot
  \bm{u}_\perp^\beta+\!
\sum_{i={3,5,6}}\!\delta^{\alpha\beta}_i\,a_i^{*} 
  \,\bm{u}_\perp^\alpha \cdot \bm{\epsilon} \cdot
  \bm{u}_\perp^\beta  \right].
\end{eqnarray}
In this expression, $\delta^{\alpha}_i\in\{0,1\}$ and
$\delta^{\alpha\beta}_i\in\{0,1\}$ are the indicatrix functions of the  
vertices of type $i$ and of the vertex pairs of type $i$, 
respectively (these types are defined in (\ref{subsubsec:type}))
, and the symbol $\sum_{<\alpha\beta>}$ indicates the
summation over all pairs $\{\alpha,\beta\}$ of vertices. 
The third line
in (\ref{Fedgeavantrans}) requires more explanations.  We
define a closed contour by grouping the vertex pairs of
types 3 (see figure~\ref{bords}), another closed contour by grouping the
vertex pairs of type 5, and three separate closed contours by
grouping the vertex pairs of type 6. The symbol $\smallosum$
denotes the sums over all these contours followed
\textit{counterclockwise}. The symbol $\eta^{\alpha\beta}$ is equal to 1
or $-1$ according to whether the direction from $\alpha$ to $\beta$ goes
outwards or inwards, respectively. Finally, in order to simplify the
notation, we have renamed $\frac{1}{2}a_{4}+\frac{1}{2}a_{8}\to a_{4}$,
$\frac{1}{2}\tilde{a}_{4}+\frac{1}{2}\tilde{a}_{8}\to \tilde{a}_{4}$,
$\frac{1}{2}\bar{a}_{4}+\frac{1}{2}\bar{a}_{8}\to \bar{a}_{4}$,
$\frac{1}{2}a_{3}^{*}+\frac{1}{2}a_{7}^{*}\to a_{3}^{*}$ and
$\frac{1}{2}\bar{\bar{a}}_{3}^{*}-\frac{1}{2}\bar{\bar{a}}_{7}^{*}\to
\bar{\bar{a}}_{3}^{*}$.

\subsection{Rotational and translational invariance}
\label{method}

Our task is now to ensure that the elastic energy remains unchanged if
the lattice is translated or rotated as a whole. Although the general
relationships are well-established (Born \& Huang 1954), it is not an easy
task to implement 
them in such a way as to put the elastic energy in a satisfactory form.
Our strategy is to rearrange $F$ in order to put forward a
number of terms that are obviously invariant, then to apply 
deliberately chosen
translations or rotations in order to show that the remaining terms
should vanish.

We first separate in $F$ ($=F_\text{bulk}+F_\text{edge}$) the
contribution $F^{(n)}$ depending on the components $u^\alpha_n$
from the contribution $F^{(\perp)}$ depending on the components
$\bm{u}^\alpha_\perp$. We thus have $F=F^{(n)}+F^{(\perp)}$ with
\begin{eqnarray}
\label{Fndebut}
F^{(n)}&=&\sum_{\alpha}\sum_{i=1}^5\delta^{\alpha}_i 
\tilde{s}_i(u^\alpha_n)^2
+\sum_{<\alpha\beta>}\sum_{i=1}^6\delta^{\alpha\beta}_i\,
  \tilde{a}_i \, u^\alpha_n u^\beta_n\,,
\\
\label{Fpdebut}
F^{(\perp)}&=&
\sum_{\alpha}\sum_{i=1}^5\delta^{\alpha}_i 
\left[s_i(\bm{u}^\alpha_\perp
)^2 
+\bar{s}_i(\bm{u}^\alpha_\perp \cdot \bm{\tau}^{\alpha})^2\right] \nonumber\\ 
&+&\!\!\sum_{<\alpha\beta>}\sum_{i=1}^6\delta^{\alpha\beta}_i\left[a_i \, 
  \bm{u}^\alpha_\perp \cdot 
  \bm{u}^\beta_\perp 
  +\bar{a}_i (\bm{u}_\perp^\alpha \cdot \bm{t}^{\alpha\beta})
  (\bm{u}_\perp^\beta \cdot \bm{t}^{\alpha\beta})\right] \nonumber \\
&+&\!\!\!\!\!\osum\!
\left[\delta^{\alpha\beta}_3\,\eta^{\alpha\beta}\,\bar{\bar{a}}_3 
  \,\bm{u}_\perp^\alpha \cdot \bm{\sigma}^{\alpha\beta} \cdot
  \bm{u}_\perp^\beta+\!
\sum_{i={3,5,6}}\!\delta^{\alpha\beta}_i\,a_i^{*} 
  \,\bm{u}_\perp^\alpha \cdot \bm{\epsilon} \cdot
  \bm{u}_\perp^\beta  \right].
\end{eqnarray}

\subsubsection{Translation invariance}

Putting forward obviously translationally invariant terms in the form of
differences, $F^{(n)}$ can be rewritten as 
\begin{equation}
\label{Fnavantrans}
F^{(n)}=-\frac{1}{2}\sum_{<\alpha\beta>}
\sum_{i={1}}^{6}
  \delta^{\alpha\beta}_i 
  \,\tilde{a}_i\,(u_n^\alpha-u_n^\beta)^2+\sum_{\alpha}\sum_{i={1}}^{5} 
  \delta^{\alpha}_i\,C_i\,(u_n^\alpha)^2,
\end{equation}
where the $C_i$ are linear combinations of the coefficients
$\tilde{s}_i$ and $\tilde{a}_j$ (the coefficients in (\ref{Fnavantrans})
still form an independent set). 
In the appendix, \S\ref{ntrans}, we show
that the invariance of $F$ under any translation along
 $\bm{n}$ requires:

\begin{equation}
\label{Ci=0}
\forall i,\ C_i=0. 
\end{equation}

Similarly, $F^{(\perp)}$ can be written as obviously translationally
invariant terms plus remainders:
\begin{eqnarray}
\label{Fpavantrans}
F^{(\perp)}&=&
-\frac{1}{2}\sum_{<\alpha\beta>}\sum_{i=1}^6
\delta^{\alpha\beta}_i\,\left\{b_i\,
(\bm{u}_\perp^\alpha-\bm{u}_\perp^\beta)^2 + \bar{b}_i
\left[(\bm{u}_\perp^\alpha-\bm{u}_\perp^\beta) \cdot
\bm{t}^{\alpha\beta}\right]^2 \right\} \nonumber 
\\ 
&-&\!\!\!\sum_{<\alpha\beta\gamma>_\text{IV}}\!\!\!
(1-\delta^{\alpha\beta\gamma}_{\text{IV},6})
\frac{2}{\sqrt{3}}\,\bar{\bar{a}}_3
\left[(\bm{u}^\beta_\perp-\bm{u}^\alpha_\perp)
\times\bm{t}^{\beta\alpha}-(\bm{u}^\beta_\perp-\bm{u}^\gamma_\perp)
\times\bm{t}^{\beta\gamma}\right]^2 \nonumber
\\
&+&\!\!\!\sum_{<\alpha\beta\gamma>_\text{III}}\,\sum_{i=3,5}
\delta^{\alpha\beta\gamma}_{\text{III},i} \,b_i^{*}\,
(\bm{u}_\perp^\alpha-\bm{u}_\perp^\gamma) \cdot \bm{\epsilon} \cdot
(\bm{u}_\perp^\beta-\bm{u}_\perp^\gamma) \nonumber 
\\
&+&\!\!\!\sum_{<\alpha\beta\gamma>_\text{V}}
\delta^{\alpha\beta\gamma}_{\text{V},6} \,b_6^{*}\,
(\bm{u}_\perp^\alpha-\bm{u}_\perp^\gamma) \cdot \bm{\epsilon} \cdot
(\bm{u}_\perp^\beta-\bm{u}_\perp^\gamma) \nonumber 
\\
&+&\!\sum_{\alpha}\sum_{i=2}^5\delta^{\alpha}_i\left[D_i\,
(\bm{u}_\perp^\alpha)^2 + \bar{D}_i\, (\bm{u}_\perp^\alpha \cdot
\bm{\tau}^\alpha)^2 \right].
\end{eqnarray}
Whereas the first line was clearly obtained by 
putting forward squares and 
discarding the remainders in the last line, the other lines deserve
detailed explanations. Here and afterwards, the symbol
$\sum_{<\alpha\beta\gamma>_\text{Y}}$ denotes the sum over all the
triangular neighbour clusters of type Y, with $\alpha$, $\beta$ and $\gamma$
taken 
in counterclockwise order. For type IV, the vertex $\beta$ is supposed
to be the central one, as in figure \ref{clusters}. The symbol
$\delta^{\alpha\beta\gamma}_{\text{Y},i}$ is the indicatrix function of
the neighbour clusters $\{\alpha,\beta,\gamma\}$ of type Y lying within the closed
contour made by the vertex pairs of type $i$ ($i=3,5$ or 6). More precisely
for $i=6$ it means to lie within at least one of the three closed
contours made by 
the vertex pairs of type 6. In the following, we shall often encounter the
translationally and 
rotationally invariant expression (as in the second
line):
\begin{equation}
\label{angle}
\left(\theta_{\alpha\beta\gamma}-\theta_{\alpha\beta\gamma}^{(0)}\right)^2
=
\left[(\bm{u}^\beta_\perp-\bm{u}^\alpha_\perp)
\times\bm{t}^{\beta\alpha}-(\bm{u}^\beta_\perp-\bm{u}^\gamma_\perp)
\times\bm{t}^{\beta\gamma}\right]^2
+\mathcal{O}(u^3)\,,
\end{equation}
where $\theta_{\alpha\beta\gamma}$ and
$\theta_{\alpha\beta\gamma}^{(0)}=\frac{\pi}{3}$ or $\frac{2\pi}{3}$ are
the angles made by the vertices triplet $\{\alpha,\beta,\gamma\}$ with 
$\beta$ being the apex, after and before an elastic distortion of the
lattice, respectively.
In the appendix, \S\,\ref{ptrans}\,(\ref{sigma}), we
show how the terms involving $\bm{\sigma}^{\alpha\beta}$ in
(\ref{Fpdebut}) were rewritten with the help of (\ref{angle}) and we
give the expressions of the coefficients $b_i$ and $\bar{b}_i$ that were
modified by this procedure. Finally, the lines involving
$\bm{\epsilon}$ were 
obtained as detailed in the appendix, \S\,\ref{ptrans}\,(\ref{epsilon}). The
coefficients $D_i$ and $\bar{D}_i$ are linear combinations of the
coefficients $s_i$, $\bar{s_i}$, $a_j$, $\bar{a}_j$, $a_j^*$ and
$\bar{\bar{a}}_3$ and the new coefficients in (\ref{Fpavantrans}) still
form an independent set.  In the appendix,
\S\,\ref{ptrans}\,(\ref{invreq}), we show that the invariance of
$F$ under in-plane translations requires:

\begin{equation}
\label{Di=0}
\forall i, \ D_i=0 \text{ and } \bar{D}_i=0.
\end{equation}

Consequently, the total energy can be written in the following form,
obviously invariant under any translation of the system:
\begin{eqnarray}
\label{Favanrot}
F&=&-\sum_{<\alpha\beta>}\sum_{i=1}^6\frac{\delta^{\alpha\beta}_i}{2}
  \left\{\tilde{a}_i(u_n^\alpha-u_n^\beta)^2 
  +b_i(\bm{u}_\perp^\alpha-\bm{u}_\perp^\beta)^2
  +\bar{b}_i\left[(\bm{u}_\perp^\alpha-\bm{u}_\perp^\beta) \cdot
  \bm{t}^{\alpha\beta}\right]^2\right\} \nonumber \\
&-&\!\!\!\sum_{<\alpha\beta\gamma>_\text{IV}}\!\!\!
(1-\delta^{\alpha\beta\gamma}_{\text{IV},6})
\frac{2}{\sqrt{3}}\,\bar{\bar{a}}_3
\left[(\bm{u}^\beta_\perp-\bm{u}^\alpha_\perp)
\times\bm{t}^{\beta\alpha}-(\bm{u}^\beta_\perp-\bm{u}^\gamma_\perp)
\times\bm{t}^{\beta\gamma}\right]^2 \nonumber
\\
&+&\!\!\!\sum_{<\alpha\beta\gamma>_\text{III}}\,\sum_{i=3,5}
\delta^{\alpha\beta\gamma}_{\text{III},i} \,b_i^{*}\,
(\bm{u}_\perp^\alpha-\bm{u}_\perp^\gamma) \cdot \bm{\epsilon} \cdot
(\bm{u}_\perp^\beta-\bm{u}_\perp^\gamma) \nonumber 
\\
&+&\!\!\!\sum_{<\alpha\beta\gamma>_\text{V}}
\delta^{\alpha\beta\gamma}_{\text{V},6} \,b_6^{*}\,
(\bm{u}_\perp^\alpha-\bm{u}_\perp^\gamma) \cdot \bm{\epsilon} \cdot
(\bm{u}_\perp^\beta-\bm{u}_\perp^\gamma)\cr
&=& F^{(n)}+F^{(\perp)},
\end{eqnarray}
where $F^{(n)}$ is the first term and $F^{(\perp)}$ is
the sum of all the other terms.

\subsubsection{Rotation invariance}

Putting forward rotationally invariant terms,
$F^{(n)}$, as obtained from (\ref{Favanrot}), can be rewritten as 
\begin{eqnarray}
\label{Fnavantrotation}
F^{(n)}&=&\frac{1}{2}k_{n 1}\!\!\!\!\!\!\!  \sum_{<\alpha\beta\gamma\delta>_\text{VI}}\!\!\!\!\!
\theta^2_{\alpha\beta\gamma\delta}
+\frac{1}{2}k_{n 2}\!\!\!\!\!\!
\sum_{<\alpha\beta\gamma\delta>_\text{VII}}\!\!\!\!\!\!
\theta^2_{\alpha\beta\gamma\delta}\nonumber
\\
&+&\!\!\!\sum_{<\alpha\beta>\,}\sum_{\,i=2,4,5,6}\delta_{\alpha\beta}^i\,
E_i\left(u_n^\alpha-u_n^\beta\right)^2+\mathcal{O}(u^3)\,.
\end{eqnarray}
The symbol $\sum_{<\alpha\beta\gamma\delta>_\text{Y}}$ denotes the sum
over all the neighbour clusters of type Y=VI or VII with the vertices placed as
shown in figure \ref{clusters}. The angle
$\theta_{\alpha\beta\gamma\delta}$ denotes the wedge angle between the
planes defined by the vertices $(\alpha,\beta,\gamma)$ and the vertices
$(\alpha,\beta,\delta)$. In the appendix,
\S\,\ref{outplanerot}\,(\ref{wedgeterms}), we show how the wedge angles are
related to the normal displacements $u_n^\alpha$. The coefficients $E_i$
are linear combinations of $\tilde{a}_i$, $\tilde{a}_1$ and
$\tilde{a}_3$, and the new coefficients still form an independent set.
We show in the appendix,
\S\,\ref{outplanerot}\,(\ref{invrotn}), that the invariance of $F$
with respect to out-of-plane rotations requires:
\begin{equation}
\label{Ei=0}
\forall i, \ E_i=0.
\end{equation}
Finally, we put forward in $F^{(\perp)}$ seven rotationally invariant
terms, obtained by considering various energy terms associated with
deviations of lengths or angles with respect to their equilibrium
values: 
\begin{eqnarray}
\label{Fperpavantrot}
F^{(\perp)}\!\!&=&\!\!
\frac{1}{2}\sum_{<\alpha\beta>}\sum_{i=1}^6\delta^{\alpha\beta}_i\,
k_i\!\left(\ell_{\alpha\beta}-\ell\right)^2
+\frac{1}{2}k_{\frac{\pi}{3}}\!\!\!\!
\sum_{<\alpha\beta\gamma>_\text{III}}\!\!\!\!
\delta^{\alpha\beta\gamma}_{\text{III},3}
\left[\left(\theta_{\alpha\beta\gamma}-
\frac{\pi}{3}\right)^2+\cp\right]\nonumber
\\
&+&\!\!\frac{1}{2}k'_{\frac{\pi}{3}}\!\!\!\!
\sum_{<\alpha\beta\gamma>_\text{V}}\!\!\!\!
\delta^{\alpha\beta\gamma}_{\text{V},6}
\left[\left(\theta_{\alpha\beta\gamma}-
\frac{\pi}{3}\right)^2+\cp\right]\nonumber
\\
&+&\!\!\frac{1}{2}k''_{\frac{\pi}{3}}\!\!\!\!
\sum_{<\alpha\beta\gamma>_\text{III}}\!\!\!\!
(\delta^{\alpha\beta\gamma}_{\text{III},5}
-\delta^{\alpha\beta\gamma}_{\text{III},3})
\left[\left(\theta_{\alpha\beta\gamma}-\frac{\pi}{3}\right)^2
+\cp\right]\nonumber
\\
&+&\!\!\frac{1}{2}k_{\frac{2\pi}{3}}\!\!\!\!
\sum_{<\alpha\beta\gamma>_\text{IV}}\!\!\!\!
\delta^{\alpha\beta\gamma}_{\text{IV},6}
\left(\theta_{\alpha\beta\gamma}-
\frac{2\pi}{3}\right)^2\nonumber
+\frac{1}{2}k'_{\frac{2\pi}{3}}\!\!\!\!
\sum_{<\alpha\beta\gamma>_\text{IV}}\!\!\!\!
\left(1-\delta^{\alpha\beta\gamma}_{\text{IV},6}\right)
\left(\theta_{\alpha\beta\gamma}-\frac{2\pi}{3}\right)^2\nonumber
\\
&+&\!\!\frac{1}{2}k_{\frac{\pi}{6}}\!\!\!\!
\sum_{<\alpha\beta\gamma>_\text{IV}}\!\!\!\!
\left(1-\delta_{\alpha\beta\gamma}^{\text{IV},6}\right) 
\left[\left(\theta_{\beta\gamma\alpha}-\frac{\pi}{6}\right)^2
\!+\left(\theta_{\beta\alpha\gamma}-\frac{\pi}{6}\right)^2\right]\nonumber
\\
&+&\!\!\!\!\!\!
\sum_{<\alpha\beta\gamma>_\text{III}}\sum_{\,i=3,5}
\delta_{\text{III},i}^{\alpha\beta\gamma}\,
G_i^*(\bm{u}_\perp^\alpha-\bm{u_\perp}^\gamma) \cdot \bm{\epsilon} \cdot
  (\bm{u}_\perp^\beta-\bm{u}_\perp^\gamma),\nonumber
\\
&+&\!\!
G_6^*\!\!\!\!\sum_{<\alpha\beta\gamma>_\text{V}}\!\!\!\!
\delta_{\text{V},6}^{\alpha\beta\gamma}\,
(\bm{u}_\perp^\alpha-\bm{u_\perp}^\gamma) \cdot \bm{\epsilon} \cdot
(\bm{u}_\perp^\beta-\bm{u}_\perp^\gamma)
+G\!\!\sum_{<\alpha\beta>}\delta^{\alpha\beta}_3
(\bm{u}^\alpha_\perp-\bm{u}^\beta_\perp)^2\,.
\end{eqnarray}
The abbreviation $\cp$ stands for `circular permutations' of
$\{\alpha,\beta,\gamma\}$ and $\ell_{\alpha\beta}$ is the
distance between the vertices $\alpha$ and $\beta$. The detail of
how (\ref{Fperpavantrot}) and (\ref{Favanrot}) are related is given in
the appendix, \S\,\ref{inplanerot}\,(\ref{Fpi}).
The coefficient $G$ is a linear combination of $b_3$ and the
other $b_i$'s 
and the coefficients $G_i^*$ are linear combinations
of the corresponding $b_i^*$ and the $b_j$'s ($j\neq 3$). The new coefficients
still form an independent set. In the appendix,
\S\,\ref{inplanerot}\,(\ref{invrotp}), we show that the invariance
of $F$ with respect to in-plane rotations implies
\begin{equation}
\label{Gi=0}
G=0 \quad \text{and} \quad \forall i, \ G^*_i=0.
\end{equation}

\subsection{Final expression of the bulk terms}
\label{final}
From the preceeding section, it follows 
that the most general elastic
energy $F$ of an hexagonal triangular lattice, invariant with respect to
translation and rotations, is given by the sum of
(\ref{Fnavantrotation}) and (\ref{Fperpavantrot}), with all the $E_i$'s,
$G_i^*$'s and G equal to zero.

Putting forward the \textit{bulk terms}, and expressing them in terms of
the displacements instead of angles and lengths (using (\ref{angle}) and the
the formulas of the appendix, \S\,\ref{outplanerot}\,(\ref{wedgeterms}) and
\S\,\ref{inplanerot}\,(\ref{Fpi})), we finally obtain
\begin{eqnarray}
\label{Fk}
F\!\!&=\!\!&
\frac{1}{2}k_{n 1}\!\!\!\!\!\!
\sum_{<\alpha\beta\gamma\delta>_\text{VI}}
\frac{4}{3\ell^2}
\left(u_n^\delta+u_n^\gamma-u_n^\alpha-u_n^\beta\right)^2 \nonumber
\\
&+&\!\!\frac{1}{2}k_{n 2}\!\!\!\!\!\!
\sum_{<\alpha\beta\gamma\delta>_\text{VII}}
\frac{4}{3\ell^2}
\left(3u_n^\alpha-u_n^\beta-u_n^\gamma-u_n^\delta\right)^2\nonumber
\\
&+&\!\!\frac{1}{2}k\!\!\sum_{<\alpha\beta>}
\frac{\delta^{\alpha\beta}_1}{\ell^2}\,
\left[(\bm{u}^\alpha-\bm{u}^\beta)\cdot\bm{t}^{\alpha\beta}\right]^2
+\frac{1}{2}k'\!\!\sum_{<\alpha\beta>}
\frac{\delta^{\alpha\beta}_2}{\ell^2}\,
\left[(\bm{u}^\alpha-\bm{u}^\beta)\cdot\bm{t}^{\alpha\beta}\right]^2\nonumber 
\\
&+&\!\!\frac{1}{2}k_{\frac{\pi}{3}}\!\!\!\!\!
\sum_{<\alpha\beta\gamma>_\text{III}}\!\!\!
\frac{1}{\ell^2}\,\delta^{\alpha\beta\gamma}_{\text{III},3} \left\{
  \left[(\bm{u}^\alpha_\perp-\bm{u}^\beta_\perp)
  \times\bm{t}^{\alpha\beta}-(\bm{u}^\alpha_\perp-\bm{u}^\gamma_\perp)
  \times\bm{t}^{\alpha\gamma}\right]^2+\cp\right\}\hspace{0.7cm}
  \nonumber \\
&+&\!\!\frac{1}{2}k_{\frac{\pi}{3}}'\!\!\!\!\!
\sum_{<\alpha\beta\gamma>_\text{V}}\!\!
\frac{1}{3\ell^2}\,\delta^{\alpha\beta\gamma}_{\text{V},6}\left\{
\left[(\bm{u}^\alpha_\perp-\bm{u}^\beta_\perp)
\times\bm{t}^{\alpha\beta}-(\bm{u}^\alpha_\perp-\bm{u}^\gamma_\perp)
\times\bm{t}^{\alpha\gamma}\right]^2+\cp\right\} \nonumber
\\
&+&\!\!\frac{1}{2}k_{\frac{2\pi}{3}}\!\!\!\!\!
\sum_{<\beta\alpha\gamma>_\text{IV}}\!\!
\delta^{\beta\alpha\gamma}_{\text{IV},6}\,\frac{1}{\ell^2}
\left[(\bm{u}^\alpha_\perp-\bm{u}^\beta_\perp)
\times\bm{t}^{\alpha\beta}-(\bm{u}^\alpha_\perp-\bm{u}^\gamma_\perp)
\times\bm{t}^{\alpha\gamma}\right]^2+\ldots
\end{eqnarray}
where $k=k_1$ and $k'=k_2$ and the ellipsis represent terms involving
only edge vertex pairs.

\section{Analysis}

\subsection{Comments on the energy form (\ref{Fk})}

\label{analysis}

\begin{figure}
\centerline{\epsfig{file=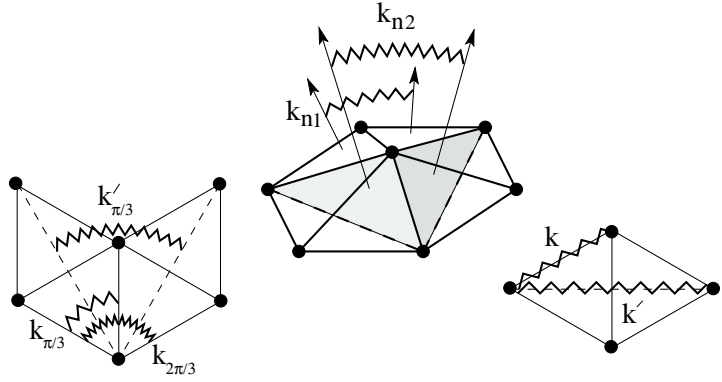,width=9.5cm}}
\caption{Interpretation of the elastic terms in (\ref{Fk}) in terms of
linear and angular springs:
The top drawing is in perspective while the
others are in projection. The arrows are normal to the facets. The
springs with elastic modulus $k$ (resp.\ $k'$) have
an equilibrium length 
$\ell$ (resp.\ $\ell\sqrt{3})$. The angular springs with moduli
$k_{n 1}$ and $k_{n 2}$ (elastic wedges) 
have zero equilibrium angles;
the angular springs with moduli $k_\psi$ or $k'_\psi$ have equilibrium
angles $\psi$.}
\label{ressorts}
\end{figure}
As shown in figure \ref{ressorts}, each term of (\ref{Fk}) can be given
a simple interpretation in terms of linear or angular springs. The terms
with coefficients $k_{n i}$, corresponding to elastic wedges,
tend to keep the lattice flat and yield a form of curvature energy. The
terms with coefficients $k$ and $k'$ represent elastic bonds acting between
nearest and next-nearest neighbours, respectively. The terms with moduli
$k_\frac{\pi}{3}$, $k'_\frac{\pi}{3}$ and $k_\frac{2\pi}{3}$ are angular springs acting at
the bond joints.

The decomposition (\ref{Fk}) is in many respects non-unique. For
instance, the term with coefficient $k_{n 2}$ could be omitted if we are only
interested in the bulk terms,
for it has the same decomposition on the bulk vertex pairs as the term
with coefficient $k_{n 1}$; this can be easily checked by comparing the
coefficients of $\delta^{\alpha\beta}_1$ and $\delta^{\alpha\beta}_2$ in
the equations of the appendix, \S\,\ref{outplanerot}\,(\ref{wedgeterms}).
Note, however, that the expressions of
these two energies in terms of clusters of four vertices differ, as we can see
in the two first terms of (\ref{Fk}).
As another example, we could have included in (\ref{Fk}) energy terms
associated with the $\frac{\pi}{6}$ angles in the bulk neighbour clusters of
type IV; 
this would bring nothing new, since they can be written as a linear
combination of other terms that are already present:
\begin{eqnarray}
&&\!\!\sum_{<\beta\alpha\gamma>_\text{IV}}\!\!\!\!
\delta^{\beta\alpha\gamma}_{\text{IV},6} 
\left[\left(\theta_{\alpha\beta\gamma}-\frac{\pi}{6}\right)^2
+\left(\theta_{\beta\gamma\alpha}-\frac{\pi}{6}\right)^2\right]
+\mathcal{O}(u^3)\nonumber
\\
&&\!\!=\frac{1}{6}\!\!\sum_{<\alpha\beta\gamma>_\text{V}}\!\!\!\!
\delta^{\alpha\beta\gamma}_{\text{V},6}
\left[\left(\theta_{\beta\alpha\gamma}-\frac{\pi}{3}\right)^2
+\cp\right]+\frac{1}{2}
\!\!\sum_{<\beta\alpha\gamma>_\text{IV}}\!\!\!\!
\delta^{\beta\alpha\gamma}_{\text{IV},6}
\left(\theta_{\beta\alpha\gamma}-\frac{2\pi}{3}\right)^2\,.\qquad
\end{eqnarray}
Nonetheless, although the \textit{form} of (\ref{Fk}) depends on the choices
we made during the various transformations on the energy, 
our procedure guarantees that it does correspond to the most general elastic
energy of a triangular lattice up to next-nearest neighbour
interactions.

\subsection{Compact expression of the elastic energy}

With the help of the formulas given in the appendix,
\S\,\ref{outplanerot}\,(\ref{wedgeterms}) and
\S\,\ref{inplanerot}\,(\ref{Fpi}), 
the energy (\ref{Fk}) can be rewritten in 
a more compact form, in terms of
the normal, transversal and longitudinal components of the relative
displacements between nearest and next-nearest neighbours :
\begin{eqnarray}
\label{FK}
F\!\!&=&\!\!\frac{1}{2}\!\!\sum_{<\alpha\beta>_\text{I}}
\!\!\left\{K_n(u_n^\alpha-u_n^\beta)^2+
K_t\left[(\bm{u}_\perp^\alpha-\bm{u}_\perp^\beta)
\cdot\bm{t}_\perp^{\alpha\beta}\right]^2
\!\!\! +K_\ell\left[(\bm{u}_\perp^\alpha-\bm{u}_\perp^\beta)
\cdot\bm{t}^{\alpha\beta}
\right]^2 \right\} \nonumber 
\\
&+&\!\!\frac{1}{2}\!\!\sum_{<\alpha\beta>_\text{II}}\!\!\!
\left\{-\frac{K_n}{3}(u_n^\alpha-u_n^\beta)^2\!+
  K_t'\left[(\bm{u}_\perp^\alpha-\bm{u}_\perp^\beta)
\cdot\bm{t}^{\alpha\beta}_\perp \right]^2 
\!\!+K_\ell'\left[(\bm{u}_\perp^\alpha-\bm{u}_\perp^\beta)
\cdot\bm{t}^{\alpha\beta}\right]^2\right\} \nonumber
\\
&+&\!\!\text{edge terms},
\end{eqnarray}
where 
$\bm{t}_\perp^{\alpha\beta}\equiv \bm{n}\times\bm{t}^{\alpha\beta}$, and
the symbols $\sum_{<\alpha\beta>_\text{I}}$ and
$\sum_{<\alpha\beta>_\text{II}}$ indicate the sum over all the neighbour
clusters 
of type I and II, i.e.\ over the pairs of nearest and next-nearest vertices,
respectively. 
The normal coefficient, which corresponds to curvature energy, is given
by $K_n=\ell^{-2}(4k_{n 1}+12k_{n 2})$. 
The transverse coefficients $K_t$ and $K_t'$ are given by
$K_t=\ell^{-2}(\frac{3}{2}k_{\pi/3}+6k_{2\pi/3})$ and
$K_t'=\ell^{-2}(\frac{1}{2}k'_{\pi/3}-\frac{3}{2}k_{2\pi/3})$, and
the
longitudinal coefficients $K_\ell$ and $K_\ell'$ are given 
by $K_\ell=\ell^{-2}(k+\frac{3}{2}k_{\pi/3})$ and
$K_\ell'=\ell^{-2}(k'+\frac{1}{2}k_{\pi/3}'+\frac{1}{2}k_{2\pi/3}).$
These new coefficients are independent linear combinations of
the former. Moreover, the five functions of
$\{\bm{u}^\alpha\}$ (proportional to the coefficients $K_n$, $K_t$,
$K_\ell$, $K'_t$ and $K'_\ell$), appearing in the form~(\ref{FK}), can be
shown to be linearly independent in their functional space. 
Therefore, the seven (or more) rigidities which one could associate with
the mechanical elements drawn in figure~\ref{ressorts} reduce to only five
independent elastic constants.

A number of comments follow. The angular coefficients of (\ref{Fk}),
i.e.\ $k_{\pi/3}$, $k_{\pi/3}'$ and $k_{2\pi/3}$, are present both in
the tranverse and the longitudinal coefficients: this is because in a
triangular lattice one cannot change the angles between bonds without changing
the distances between the vertices. Conversely, the coefficients $k$ and
$k'$ appear only in the longitudinal coefficients. Also, as explained
above, the two coefficients $k_{n 1}$ and $k_{n 2}$ collapse into a
unique normal coefficient $K_n$. 
Most importantly, note that the non-central term
with coefficients $K_t$, which was understood from Keating (1966) to violate
rotational invariance, is actually allowed (even if we restrain the model to
nearest neighbour interaction only). We shall comment on this in
\S\ref{conclusion}.

Finally, although (\ref{FK}) is expressed as a sum of squared terms, the
stability issue is not trivial, since these terms are not independent. 
For instance, the stability of the undistorted lattice does not require
all the five coefficients of (\ref{FK}) to be positive. 
For example, if all the coefficients in (\ref{Fk}) are positive, which
is obviously a sufficient condition for stability, and if
$k_\frac{2\pi}{3}>\frac{1}{3}k'_{\pi/3}$, we may have  $K'_t<0$. Further
comments on the statibility issue will be given in \S4.

\subsection{Link with the continuous theory}

In a continuous description, the displacement of the point
$\bm{r}$ on the plane $\mathcal{P}$ is described by a three-dimensional
continuous function $\bm{u}(\bm{r})$. In order to link the
discrete description to the continuous one, 
we assume that the displacements $\bm{u}^\alpha$
vary very slowly at the scale of the lattice spacing $\ell$ and we
identify $\bm{u}^\alpha$ and $\bm{u}(\bm{r}^\alpha)$. Hence, we
transform (\ref{FK}) with the help of the Taylor expansion:
\begin{equation}
\bm{u}^\beta-\bm{u}^\alpha=\xi\,\bm{t}^{\alpha\beta}\cdot\bm{\nabla}\bm{u}
\left(\bm{r}^\alpha\right)
+\mathcal{O}(\xi^2)\,,
\end{equation}
where $\bm{r}^\beta-\bm{r}^\alpha=\xi\,\bm{t}^{\alpha\beta}$, with
$\xi=\ell$ for nearest neighbour vertices and $\xi=\ell\sqrt{3}$ for
next-nearest neighbour vertices, and
$\sum_\alpha\to2\ell^{-2}/\sqrt{3}\int\!\rd^2r$.
We take an orthonormal coordinate system $(x,y,z)$, with the $z$ axis parallel to
$\bm{n}$ and the $x$ axis along one of the three nearest-neighbour directions
of the undistorted lattice. Then, for the in-plane part $F^{(\perp)}$ of
(\ref{FK}), we obtain $F^{(\perp)}=\int\!\rd^2r\,f^{(\perp)}$, with
\begin{eqnarray}
f^{(\perp)}\!\!\!&=&\!\!\!\bar{K}_\ell
\left[3\left(u^x_{,x}\right)^2+3\left(u^y_{,y}\right)^2
+\left(u^x_{,y}\right)^2+\left(u^y_{,x}\right)^2 
+2\,u^x_{,x}\,u^y_{,y}+2\,u^x_{,y}\,u^y_{,x}\right]\nonumber
\\
&+&\!\!\!\bar{K}_t
\left[\left(u^x_{,x}\right)^2
+\left(u^y_{,y}\right)^2
+3\left(u^x_{,y}\right)^2+3\left(u^y_{,x}\right)^2
-2\,u^x_{,x}\,u^y_{,y}-2\,u^x_{,y}\,u^y_{,x}\right],\quad
\end{eqnarray}
in which $\bar{K}_\ell=\frac{1}{8}\sqrt{3}(K_\ell+3K'_\ell)$ and
$\bar{K}_t=\frac{1}{8}\sqrt{3}(K_t+3K'_t)$ and
$u_{,j}^i$ is the derivative of the $i$th component of $\bm{u}$
with respect to the $j$th component of $\bm{r}$ (note that
$i\in\{x,y,z\}$ whereas $j\in\{x,y\}$).
Using integrations by parts, we can rewrite the energy density in the
standard, isotropic, form
$f^{(\perp)}=\frac{1}{2}\lambda\,e_{ii}\,e_{jj}+\mu\,e_{ij}\,e_{ij}$,
with $e_{ij}=\frac{1}{2}(u^i_{,j}+u^j_{,i})$. The Lam\'e coefficients
are given by
\begin{eqnarray}
\lambda\!\!&=&\!\!\frac{\sqrt{3}}{4}\left[K_\ell+3K_\ell'
-5\left(K_t+3K_t'\right)\right],
\\
\mu\!\!&=&\!\!\frac{\sqrt{3}}{4}\left[K_\ell+3K_\ell'
+3\left(K_t+3K_t'\right)\right].
\end{eqnarray}
In terms of the spring constants of (\ref{Fk}), this gives
$\lambda=\frac{3}{2}\sqrt{3}\,\ell^{-2}(\frac{1}{6}k+\frac{1}{2}k'
-k_{\pi/3}-k'_{\pi/3}-k_{2\pi/3})$ and
$\mu=\frac{3}{2}\sqrt{3}\,\ell^{-2}(\frac{1}{6}k+\frac{1}{2}k'
+k_{\pi/3}+k'_{\pi/3}+k_{2\pi/3})$. For a system composed only of
springs between nearest neighbours, we recover the classical relation
$\lambda=\mu=\frac{1}{4}\sqrt{3}\,k/\ell^2$ (Kantor \textit{et al.} 1987). As
expected, the area compressibility modulus $K_A=\lambda+\mu$
depends only on the spring constants and not on the angular constants:
 $\lambda+\mu=\frac{1}{2}\sqrt{3}\,(K_\ell+3K_\ell'-K_t-3K_t')
=\frac{1}{2}\sqrt{3}\,\ell^{-2}(k+3k')$.

For the normal part $F^{(n)}$ of the energy, it is necessary to add the
second order terms $\frac{1}{2}\xi^2\,\bm{t}^{\alpha\beta}\cdot
\bm{\nabla}\bm{\nabla}\bm{u} \cdot\bm{t}^{\alpha\beta}$ in the Taylor
expansion because the first order vanishes. With
$F^{(n)}=\int\!\rd^2r\,f^{(n)}$, we obtain the isotropic energy density
\begin{equation}
f^{(n)}=\frac{1}{2}\kappa \left(u^z_{,xx}+u^z_{,yy}\right)^2,
\end{equation}
with the bending constant $\kappa=\frac{1}{8}\ell^2\sqrt{3}\,K_n
=\frac{\sqrt{3}}{2}(k_{n 1}+3k_{n 2})$. 
One recognises the bending energy associated with the curvature
$H=\frac{1}{2}(u^z_{,xx}+u^z_{,yy})+\mathcal{O}(u^3)$ of the surface
$z=u^z(\bm{r})$.

\section{Summary and discussion}
\label{conclusion}

      In this paper, we have determined the most general form for the
elastic energy of a triangular lattice when both first and second
neighbour couplings are taken into account.
        We have found that, among the six coefficients originating from the
decomposition of the energy in terms of the normal, transversal and
longitudinal components of the relative displacements between nearest
and next-nearest neighbours, only five are independent (equation (3.2)).
Consequently, only five combinations of the elastic parameters given for
instance in figure~\ref{ressorts} are measurable.
        We have also found that only three coefficients survive in the large
scale limit, the Lam\'e coefficients and the bending modulus; therefore
our discrete description introduces two new elastic coefficients.
From the following well-known stability criteria: $\kappa>0$, $\mu>0$
and $\lambda+\mu>0$, it is possible to deduce some necessary stability
conditions: $K_n>0$, $K_\ell+3K_\ell'>0$ and
$K_t+3K_t'>-\frac{1}{3}K_\ell-K_\ell'$. A complete stability analysis
can be performed by Fourier transform. However, the additional
stability conditions on the coefficients being rather intricate, we
shall not describe them here.

        One of the main issue in the paper was the necessity to take into
account the boundary elastic energy in order to correctly impose
rotational invariance. This is well exemplified by the following energy
term, which represents the sum of the energies of the angular springs
between nearest bonds:
\begin{eqnarray}
\sum_i\left(\theta_i-\frac{\pi}{3}\right)^2\!\!\!&=&\!\!\!
\sum_{\alpha-\beta}^{(B)}
\frac{3}{2}\left(\bm{u}^\alpha_\perp-\bm{u}^\beta_\perp\right)^2\nonumber\\
&+&\!\!\!\sum_{\alpha-\beta}^{(\partial B)}
\frac{3}{4}\left(\bm{u}^\alpha_\perp-\bm{u}^\beta_\perp\right)^2
+\osumtrait\frac{3\sqrt{3}}{2}\,
\bm{u}_\perp^\alpha\cdot\bm{\epsilon}\cdot\left(\bm{u}^\beta_\perp
-\bm{u}^\alpha_\perp\right),
\end{eqnarray}
where $i$ runs over all the pairs of adjacent bonds, making an angle
$\theta_i$, and $\alpha\!-\!\beta$ runs over the interior bonds when $(B)$
is indicated, or over the boundary bonds when $(\partial B)$ is
indicated. In the last sum, the bonds ($\alpha\!\!\to\!\!\beta$) must be
taken counterclockwise. Indeed, the left-hand side is manifestly
rotationally invariant and, under an infinitesimal in-plane rotation
$\bm{u}^\alpha_\perp=\omega\,\bm{n}\times\bm{r}^\alpha$, the last term
of the right-hand side (r.-h.~s.), which is a boundary term, can be
shown to yield an extensive contribution $-2\omega^2\,A$, where $A$ is
the area of the lattice, just as the first term of the r.-h.~s, which is
a bulk term. It follows that: i) this first term is not rotationally
invariant by itself, ii) that boundary terms are therefore necessary to
correctly ensure rotational invariance, and that, as announced, iii) the
term with coefficient $K_t$ is indeed allowed, even if we restrain the
model to nearest neighbour interaction only.

In conclusion, the method developed in this paper---which consists in
defining neighbour clusters based on the interaction range, in
considering explicitely the edges of the lattice, in considering the
local symmetries defined in terms of cluster environment to distinguish
the various edge terms, etc.---provides a safe basis to further
implement rotational invariance.  This method could be applied to a
large variety of lattices and to higher dimensions. 
\appendix{Translational and rotational invariances}

This appendix contains the detail of the calculations made to impose
translational and rotational invariances to the elastic energy.

\subsection{Translation parallel to $\bm{n}$}
\label{ntrans}
Here we demonstrate (\ref{Ci=0}), on the basis of (\ref{Fnavantrans}).
Substituting $\bm{u}^\alpha=\xi\,\bm{n}$ in equation
(\ref{Fnavantrans}) yields $F^{(n)}/\xi^2 =\sum_{i=1}^5C_iV_i$, where the
coefficients $V_i$ are defined in the table of figure~\ref{tablo}. 
Hence, with $\xi=1$,
\begin{eqnarray}
\label{ftransnorm}
F^{(n)}=
3N^2 C_1+3N\left(-9C_1+6C_2+6C_4\right)+7C_1-12C_2+6C_3-6C_4+6C_5\,\,\,
\end{eqnarray}
Setting $F^{(n)}=0$ for all $N$ implies $0=C_1=C_2+C_4=C_3+C_4+C_5$.
Next, imposing the same translation but from a
\textit{distorted} state, defined by $\bm{u}^\alpha=\bm{0}$
except for one vertex $\alpha_0$ of type 4 (resp. 5) for which
$\bm{u}^{\alpha_0}=\xi_0 \bm{n}$, it is straightforward to show that the
invariance of $F$ requires $C_4=0$ (resp. $C_5=0$). Consequently, all
the coefficients $C_i$ must vanish.

\begin{figure}
\begin{center}
\begin{tabular}{|c||c|c|}
\hline
Type & Vertex & Pairs of vertices \\
\hline\hline
1 & $V_1=3N^2-9N+7$ & $P_1=9N^2-15N+6 $ \\
\hline
2 & $V_2=6N-12$ & $P_2=9N^2-15N+6$\\
\hline
3 & $V_3=6$    & $P_3=12N-12$ \\
\hline
4 & $V_4=6N-6$  & $P_4=6$  \\
\hline
5 & $V_5=6$      & $P_5=6N$ \\
\hline
6 &         & $P_6=12N-6$\\
\hline 
\end{tabular}
\end{center}
\caption{Number of vertices, $V_i$, and of pair of vertices, $P_i$, as a
function of their type (see figure \ref{bords}) for a regular hexagon
of edge length $N\ell$.}
\label{tablo}
\end{figure}

\subsection{Translation parallel to $\mathcal{P}$}
\label{ptrans}
Here we display in (\ref{sigma}) and (\ref{epsilon}) some
details of the calculations made to go from (\ref{Fpdebut}) to
(\ref{Fpavantrans}) and we demonstrate in (\ref{invreq}) the equation
(\ref{Di=0}) on the basis of equation (\ref{Fpavantrans}).

\subsubsection{Treatment of the terms involving $\bm{\sigma}^{\alpha\beta}$.}
\label{sigma}
In the process to obtain equation (\ref{Fpavantrans}), the terms
involving 
$\bm{\sigma}^{\alpha\beta}$ in 
(\ref{Fpdebut}) have been eliminated
using the following identity:
\begin{eqnarray}
\label{transsigma}
&&\!\!\!\!\!\!
\sum_{<\alpha\beta\gamma>_\text{IV}}\!\!\!\!\!
\left(1-\delta^{\alpha\beta\gamma}_{\text{IV},6}\right)\frac{2}{\sqrt{3}}
  \left[(\bm{u}^\beta_\perp-\bm{u}^\alpha_\perp)
  \times\bm{t}^{\beta\alpha}-(\bm{u}^\beta_\perp-\bm{u}^\gamma_\perp)
  \times\bm{t}^{\beta\gamma}\right]^2 \nonumber 
\\
&&\!\!\!\!\!=\!\!\!\!\sum_{<\alpha\beta\gamma>_\text{IV}}\!\!\!\!\!
\left(1-\delta^{\alpha\beta\gamma}_{\text{IV},6}\right)\frac{8}{3\sqrt{3}}
  \left[(\bm{u}^\alpha_\perp-\bm{u}^\beta_\perp)\!
  \cdot\!\left(\bm{t}^{\beta\gamma}\!+\!\frac{\bm{t}^{\beta\alpha}}{2}\right)
  +(\bm{u}^\gamma_\perp-\bm{u}^\beta_\perp) 
  \!\cdot\!\left(\bm{t}^{\beta\alpha}\!+\!\frac{\bm{t}^{\beta\gamma}}{2}\right)
\right]^2\!\! \nonumber 
\\
&&\!\!\!\!\!=\sum_\alpha\frac{\sqrt{3}}{2}\,\left(\delta^{\alpha}_2
  -\delta^{\alpha}_3+\delta^{\alpha}_4\right)
  \left[\left(\bm{u}_\perp^\alpha\right)^2
-2\left(\bm{u}^\alpha\cdot\bm{\tau}^\alpha\right)^2\right] \nonumber 
\\
&&\!\!\!\!\!-\osum\left[\delta^{\alpha\beta}_3\,\eta^{\alpha\beta}\, 
\bm{u}^\alpha\cdot\bm{\sigma}^{\alpha\beta}\cdot\bm{u}^\beta
+\left(\delta^{\alpha\beta}_3+2\delta^{\alpha\beta}_5
-\delta^{\alpha\beta}_6\right)
\left(\bm{u}^\alpha\cdot\bm{\epsilon}\cdot\bm{u}^\beta\right)\right] \nonumber
\\
&&\!\!\!\!\!+\sum_{<\alpha\beta>}\!\!\!\sqrt{3}\,
\bigg\{\!\!\left(\delta^{\alpha\beta}_3
  +2\delta^{\alpha\beta}_5-\frac{1}{2}\delta^{\alpha\beta}_6\right)
  \left(\bm{u}_\perp^\alpha-\bm{u}_\perp^\beta\right)^2  \nonumber 
\\
&&\qquad\qquad\left.+\left(-\delta^{\alpha\beta}_3-2\delta^{\alpha\beta}_5 
  +\frac{2}{3}\delta^{\alpha\beta}_6 \right)
  \left[\left(\bm{u}^\alpha-\bm{u}^\beta\right)
\cdot\bm{t}^{\alpha\beta}\right]^2\right\}.
\end{eqnarray}
The coefficients $b_i$ and $\bar{b}_i$ in equation (\ref{Fpavantrans})
are then given by 
$b_1=a_1$, $\bar{b}_1=\bar{a}_1$, $b_2=a_2$, $\bar{b}_2=\bar{a}_2$,
$b_3=a_3-2\sqrt{3}\,\bar{\bar{a}}_3$,
$\bar{b}_3=\bar{a}_3+2\sqrt{3}\,\bar{\bar{a}}_3$,
$b_4=a_4$, $\bar{b}_4=\bar{a}_4$, $b_5=a_5-4\sqrt{3}\,\bar{\bar{a}}_3$,
$\bar{b}_5=\bar{a}_5+4\sqrt{3}\,\bar{\bar{a}}_3$,
$b_6=a_6+\sqrt{3}\,\bar{\bar{a}}_3$,
$\bar{b}_6=\bar{a}_6-\frac{4}{3}\sqrt{3}\,\bar{\bar{a}}_3$,
$b_3^*=a_3^*+\bar{\bar{a}}_3$, $b_5^*=a_5^*+2\bar{\bar{a}}_3$,
$b_6^*=a_6^*-\bar{\bar{a}}_3$.

\subsubsection{Discrete Stokes-like formula: 
treatment of the terms involving $\bm{\epsilon}$}
\label{epsilon}
In the process of the transformation from (\ref{Fpdebut})
to (\ref{Fpavantrans}), the terms involving $\bm{\epsilon}$
have been rewritten
using the following general Stokes-like
theorem transforming a sum of edge terms into a sum of bulk terms:
\begin{equation}
\osum\delta^{\alpha\beta}_i\,
  \bm{u}_\perp^\alpha \cdot \bm{\epsilon}^{\alpha\beta} \cdot
  \bm{u}_\perp^\beta\,=\!\!\!
  \sum_{<\alpha\beta\gamma>_\text{Y}}
\!\!\!\delta^{\alpha\beta\gamma}_{\text{Y},i}\,
  (\bm{u}_\perp^\alpha-\bm{u_\perp}^\gamma) \cdot \bm{\epsilon} \cdot
  (\bm{u}_\perp^\beta-\bm{u}_\perp^\gamma)\,,
\end{equation}
where Y=III for $i=3,5$ and Y=V for $i=6$, and where the vertices $\alpha$,
$\beta$ and $\gamma$ are taken in counterclockwise order in the right-hand
side of the equation.

\subsubsection{Invariance requirements}
\label{invreq}
Here and afterwards, $\bm{t}$ will denote any of the unitary vectors
joining two nearest-neighbours vertices of the undistorted lattice. First we
substitute 
$\bm{u}^\alpha=\xi\,\bm{t}$ in equation (\ref{Fpavantrans}), which yields a
third-order polynomial in $N$.  Setting $F=0$ for all $N$ implies
$0=D_1=2D_2+\bar{D}_2+2D_4+\bar{D}_4=
2D_3+\bar{D}_3+2D_4+\bar{D}_4+2D_5+\bar{D}_5$, which leaves six
independent coefficients among the $D_i$ and $\bar{D}_i$. 
Then, by imposing the same translation but on properly chosen
distorted states, it 
is easy to show 
that all the coefficients $D_i$ and $\bar{D}_i$ must be set equal to
zero to satisfy the translational invariance.

\subsection{Rotation with the rotation axis parallel to $\mathcal{P}$}
\label{outplanerot}

In this paragraph, we (\ref{wedgeterms}) display some
details of the calculations made to go from (\ref{Favanrot}) to
(\ref{Fnavantrotation}) and (\ref{invrotn}) demonstrate the equation
(\ref{Ei=0}) on the basis of equation (\ref{Fnavantrotation}).

\subsubsection{Decomposition of the terms involving wedge angles}
\label{wedgeterms}

\label{Fni}
Expanding the first and second terms in (\ref{Fnavantrotation}) yields 
\begin{eqnarray}
\label{Fn1}
&&\sum_{<\alpha\beta\gamma\delta>_\text{VI}}\!\!\!\!\!
\theta^2_{\alpha\beta\gamma\delta}
=\!\!\!\!\!
\sum_{<\alpha\beta\gamma\delta>_\text{VI}}\!\!\!
 \frac{4}{3\ell^2}\left(u_n^\alpha+u_n^\beta-u_n^\gamma-u_n^\delta\right)^2
+\mathcal{O}(u^3)\ \nonumber
\\
&&=\frac{4}{3\ell^2}\!\!\sum_{<\alpha\beta>}\!\!\left(
3\delta^{\alpha\beta}_1
-\delta^{\alpha\beta}_2+2\delta^{\alpha\beta}_3
+\delta^{\alpha\beta}_4+2\delta^{\alpha\beta}_5
-\delta^{\alpha\beta}_6
\right)\left(u_n^\alpha-u_n^\beta\right)^2\!+\mathcal{O}(u^3),\qquad\,\,
\end{eqnarray}
and
\begin{eqnarray}
\label{Fn2}
&&\sum_{<\alpha\beta\gamma\delta>_\text{VII}}\!\!\!\!\!
\theta^2_{\alpha\beta\gamma\delta}
=\!\!\!\!\!
\sum_{<\alpha\beta\gamma\delta>_\text{VII}}\!\!
\frac{4}{3\ell^2}\left(3u_n^\alpha-u_n^\beta-u_n^\gamma-u_n^\delta\right)^2
+\mathcal{O}(u^3) \nonumber 
\\
&&=\frac{4}{\ell^2}\!\!\sum_{<\alpha\beta>}\left(6\delta^{\alpha\beta}_1
-2\delta^{\alpha\beta}_2+3\delta^{\alpha\beta}_3
+3\delta^{\alpha\beta}_4
-\delta^{\alpha\beta}_6\right)\left(u_n^\alpha-u_n^\beta\right)^2
+\mathcal{O}(u^3).\qquad
\end{eqnarray}
The coefficients $k_{n i}$ are then given by
$k_{n 1}=\frac{3}{4}\ell^2(\tilde{a}_1-2\tilde{a}_3)$ and
$k_{n 2}=-\frac{1}{12}\ell^2(2\tilde{a}_1-3\tilde{a}_3)$.

\subsubsection{Invariance requirements}
\label{invrotn}

In an infinitesimal rotation with the rotation axis parallel to one of
the directions of the undistorted lattice, we have
$\bm{u}^\alpha=\theta\,\bm{t}\times\bm{r}^\alpha$, 
which is parallel to $\bm{n}$.
Hence $\bm{u}^\alpha-\bm{u}^\beta=-\theta\ell\, \bm{t} \times
\bm{t}^{\alpha\beta}$ for nearest neighbours, and $
\bm{u}^\alpha-\bm{u}^\beta=-\theta\ell\sqrt{3}\, \bm{t} \times
\bm{t}^{\alpha\beta}$ for next-nearest neighbours.  Substituting these
expressions into (\ref{Fnavantrotation}) yields
$F^{(n)}/(\frac{1}{2}\theta^2\ell^2)=3E_2P_2+E_4P_4+E_5P_5+3E_6P_6$,
where the coefficients $P_i$ are given in table~\ref{tablo}.  Setting
$F^{(n)}=0$ for all $N$ implies $E_2=0$, $E_5=-6E_6$ and $E_4=3E_6$,
which leaves only one independent coefficient. Next, imposing the same
rotation but on a properly chosen distorted state, it can easily
been shown that the last $E_i$ must be set equal to zero to satisfy the 
rotational invariance.

\subsection{Rotation within the plane $\mathcal{P}$}
\label{inplanerot}

In this paragraph, we (\ref{Fpi}) display some
details of the calculations made to go from (\ref{Favanrot}) to
(\ref{Fperpavantrot}) and (\ref{invrotp}) demonstrate the equation
(\ref{Gi=0}) on the basis of equation (\ref{Fperpavantrot}).

\subsubsection{Expansion of the various terms in (\ref{Fperpavantrot})}
\label{Fpi}

The first term in (\ref{Fperpavantrot}) is expanded as
follows:
\begin{equation}
\sum_{<\alpha\beta>}\sum_{i=1}^6
\delta^{\alpha\beta}_i\,k_i\left(\ell_{\alpha\beta}-\ell\right)^2
=\sum_{<\alpha\beta>}\sum_{i=1}^6
\delta^{\alpha\beta}_i\frac{k_i}{\ell^2}\left[(\bm{u}_\perp^\alpha
-\bm{u}_\perp^\beta)\cdot\bm{t}^{\alpha\beta}\right]^2+\mathcal{O}(u^3).
\end{equation}
Setting $k_{2\pi/3}=\ell^2(-2b_2+4b_6-\frac{1}{3}b_4+\frac{2}{3}b_5)$,
the relations between the coefficients $k_i$ and the coefficients in
(\ref{Favanrot}) are found to be $k_1=\ell^2(-\bar{b}_1+6k_{2\pi/3})$,
$k_2=-\ell^2(\bar{b}_2+2k_{2\pi/3})$,
$k_3=\ell^2(-\bar{b}_3+3k_{2\pi/3})$,
$k_4=\ell^2(-\bar{b}_4+3k_{2\pi/3})$, $k_5=-\ell^2\bar{b}_5$,
$k_6=-\ell^2(\bar{b}_6+k_{2\pi/3})$.

The terms with coefficients $k_\frac{\pi}{3}$, $k'_\frac{\pi}{3}$ and
$k''_\frac{\pi}{3}$ in (\ref{Fperpavantrot}) can be expanded using
the identity (\ref{angle}) and the following identity:
\begin{eqnarray}
&&\sum_{<\alpha\beta\gamma>_\text{Y}}\!\!\!\!\!
\Delta^{\alpha\beta\gamma}
\left\{\left[(\bm{u}^\beta_\perp-\bm{u}^\alpha_\perp)
\times\bm{t}^{\beta\alpha}-(\bm{u}^\beta_\perp-\bm{u}^\gamma_\perp)
\times\bm{t}^{\beta\gamma}\right]^2\!\!+\cp\right\}\nonumber
\\
&&=\frac{3}{2}\!\sum_{<\alpha\beta>}\!\!
c^{\alpha\beta}\,
(\bm{u}^\alpha_\perp
-\bm{u}^\beta_\perp)^2 
+\frac{3\sqrt{3}}{2}\!\!\!\!\!\sum_{<\alpha\beta\gamma>_\text{Y}}\!\!\!\!\!
\Delta^{\alpha\beta\gamma}
(\bm{u}_\perp^\alpha-\bm{u}_\perp^\gamma) \cdot
\bm{\epsilon} \cdot
(\bm{u}_\perp^\beta-\bm{u}_\perp^\gamma),\qquad\,\,\,
\end{eqnarray}
where either $\text{Y}=\text{III}$,
$\Delta^{\alpha\beta\gamma}=\delta^{\alpha\beta\gamma}_{\text{III},3}$,
$c^{\alpha\beta}=\delta^{\alpha\beta}_1+\frac{1}{2}\delta^{\alpha\beta}_3$,
or $\text{Y}=\text{V}$,
$\Delta^{\alpha\beta\gamma}=\delta^{\alpha\beta\gamma}_{\text{V},6}$,
$c_{\alpha\beta}=\delta^{\alpha\beta}_2+\frac{1}{2}\delta^{\alpha\beta}_6$,
or $\text{Y}=\text{III}$,
$\Delta^{\alpha\beta\gamma}=\delta^{\alpha\beta\gamma}_{\text{III},5}
-\delta^{\alpha\beta\gamma}_{\text{III},3}$,
$c_{\alpha\beta}=\frac{1}{2}\delta^{\alpha\beta}_3
+\delta^{\alpha\beta}_4+\frac{1}{2}\delta^{\alpha\beta}_5$.
It follows that the relationships between the coefficients are
$k_{\pi/3}=\ell^2(-\frac{2}{3}b_1+8b_2-16b_6+\frac{4}{3}b_4
-\frac{8}{3}b_5)$, $k'_{\pi/3}=\ell^2(-b_4+2b_5-8b_2+12b_6)$,
$k''_{\pi/3}=\ell^2(4b_2-8b_6-\frac{4}{3}b_5)$.

The terms with the coefficient $k_\frac{2\pi}{3}$
in (\ref{Fperpavantrot}) can be expanded using the identity
(\ref{angle}) and the following identity:
\begin{eqnarray}
&&\sum_{<\alpha\beta\gamma>_\text{IV}}\!\!\!\!\!
\Delta'^{\alpha\beta\gamma}
\left[(\bm{u}^\beta_\perp-\bm{u}^\alpha_\perp)
\times\bm{t}^{\beta\alpha}-(\bm{u}^\beta_\perp-\bm{u}^\gamma_\perp)
\times\bm{t}^{\beta\gamma}\right]^2\nonumber
\\
&&~~~~~~~~=3\!\!\sum_{<\alpha\beta>}\left\{
d^{\alpha\beta}\,
(\bm{u}^\alpha_\perp -\bm{u}^\beta_\perp)^2
+d'^{\alpha\beta}\left[
(\bm{u}^\alpha_\perp -\bm{u}^\beta_\perp)\cdot\bm{t}^{\alpha\beta}
\right]^2
\right\}\qquad\nonumber
\\
&&~~~~~~~~+\frac{\sqrt{3}}{2}\!
\sum_{<\alpha\beta\gamma>_\text{V}}\!\!\!\!
\Delta'^{\alpha\beta\gamma}\,
(\bm{u}_\perp^\alpha-\bm{u}_\perp^\gamma) \cdot
\bm{\epsilon} \cdot (\bm{u}_\perp^\beta-\bm{u}_\perp^\gamma),
\end{eqnarray}
where
$\Delta'^{\alpha\beta\gamma}=\delta^{\alpha\beta\gamma}_{\text{IV},6}$,
$d^{\alpha\beta}=2\delta^{\alpha\beta}_1-\frac{1}{2}\delta^{\alpha\beta}_2+
\delta^{\alpha\beta}_3+\delta^{\alpha\beta}_4
-\frac{1}{4}\delta^{\alpha\beta}_6$,
$d'^{\alpha\beta}=-2\delta^{\alpha\beta}_1+\frac{2}{3}\delta^{\alpha\beta}_2
-\delta^{\alpha\beta}_3-\delta^{\alpha\beta}_4
+\frac{1}{3}\delta^{\alpha\beta}_6$.

The terms with coefficient $k'_\frac{2\pi}{3}$
in (\ref{Fperpavantrot}) can be expanded using the identities
(\ref{angle}) and (\ref{transsigma}). It follows that
$k'_{2\pi/3}=\ell^2(2b_2-4b_6-\frac{4}{\sqrt{3}}\bar{\bar{a}}_3)$.

Finally, the term with coefficient $k_\frac{\pi}{6}$ in
(\ref{Fperpavantrot}) can be expanded using 
\begin{eqnarray}
&&\sum_{<\beta\alpha\gamma>_\text{IV}}\!\!
\left(1-\delta^{\beta\alpha\gamma}_{\text{IV},6}\right)
  \left\{\left[(\bm{u}^\beta_\perp-\bm{u}^\alpha_\perp)
  \times\bm{t}^{\beta\alpha}-\frac{1}{\sqrt{3}}\,
  (\bm{u}^\beta_\perp-\bm{u}^\gamma_\perp) 
  \times\bm{t}^{\beta\gamma}\right]^2 \right.\nonumber 
\\
&&\qquad\qquad\qquad\qquad\,\,\left.
  +\left[(\bm{u}^\gamma_\perp-\bm{u}^\alpha_\perp) 
  \times\bm{t}^{\gamma\alpha}-\frac{1}{\sqrt{3}}\,
  (\bm{u}^\gamma_\perp-\bm{u}^\beta_\perp) 
  \times\bm{t}^{\gamma\beta}\right]^2\right\}\nonumber
\\
&&~~~~~~=\sum_{<\alpha\beta>}\left\{
e^{\alpha\beta}\,
(\bm{u}^\alpha_\perp -\bm{u}^\beta_\perp)^2
+e'^{\alpha\beta}\left[
(\bm{u}^\alpha_\perp -\bm{u}^\beta_\perp)\cdot\bm{t}^{\alpha\beta}
\right]^2
\right\}\qquad\nonumber
\end{eqnarray}
\begin{eqnarray}
&&~~~~~~+\frac{1}{\sqrt{3}}\!
\sum_{<\alpha\beta\gamma>_\text{III}}
\!\!\!\!
\left(\delta^{\alpha\beta\gamma}_{\text{III},3}+
2\delta^{\alpha\beta\gamma}_{\text{III},5}\right)
(\bm{u}_\perp^\alpha-\bm{u}_\perp^\gamma) \cdot
\bm{\epsilon} \cdot (\bm{u}_\perp^\beta-\bm{u}_\perp^\gamma)\nonumber
\\
&&~~~~~~-\frac{1}{\sqrt{3}}\!
\sum_{<\alpha\beta\gamma>_\text{V}}
\!\!\!\!
\delta^{\alpha\beta\gamma}_{\text{V},6}\,
(\bm{u}_\perp^\alpha-\bm{u}_\perp^\gamma) \cdot
\bm{\epsilon} \cdot (\bm{u}_\perp^\beta-\bm{u}_\perp^\gamma).
\end{eqnarray}
The supplementary factor $1/\sqrt{3}$ which does not appear in (\ref{angle}) 
is due to the factor $\sqrt{3}$ between the
equilibrium distances $\beta-\alpha$ and $\beta-\gamma$ (where $\beta$
is the apex of the angle).  The coefficients of the expansion are
$e^{\alpha\beta}=\frac{3}{4}\delta^{\alpha\beta}_3
+\frac{3}{2}\delta^{\alpha\beta}_5-\frac{1}{3}\delta^{\alpha\beta}_6$,
$e'^{\alpha\beta}=-\frac{1}{2}\delta^{\alpha\beta}_3-\delta^{\alpha\beta}_5
+\frac{1}{3}\delta^{\alpha\beta}_6$, and $k_{\pi/6}=\ell^2(-6b_2+12b_6)$.

\subsubsection{Invariance requirements}
\label{invrotp}
An infinitesimal rotation around an axis parallel to $\bm{n}$ of the
undistorted lattice yields
$\bm{u}^\alpha=\theta\bm{n}\times\bm{r}^\alpha$, parallel to
$\mathcal{P}$. Hence, $\bm{u}^\alpha-\bm{u}^\beta=
-\theta\ell\,\bm{n}\times\bm{t}^{\alpha\beta}$ for nearest neighbours,
and $\bm{u}^\alpha-\bm{u}^\beta=-\theta\ell\sqrt{3}\,\bm{n}\times
\bm{t}^{\alpha\beta}$ for next-nearest neighbours.
Substituting these expressions into (\ref{Fperpavantrot}) yields
$F^{(\perp)}/(\theta^2\ell^2)=-\frac{3}{2}\sqrt{3}N^2(G_3^*+G_5^*+3G_6^*)
+3N(4G+\sqrt{3}\,G_3^*+3\sqrt{3}\,G_6^*)-3(4G+\sqrt{3}\,G_6^*)$.
Setting $F^{(\perp)}=0$ for all $N$ gives $\sqrt{3}\,G_3^*=8G$,
$\sqrt{3}\,G_5^*=4G$ and $\sqrt{3}\,G_6^*=-4G$. Finally, imposing the same
rotation on a properly chosen initial state allows to deduce that
$G$ must be set equal to zero to satisfy the rotational invariance.

\end{document}